\documentclass[epsfig]{kluwer}
\usepackage{graphicx}
\newdisplay{Matteucci}{What determines galactic evolution?}

\begin{document}
\begin{article}
\begin{opening}

\title{What determines galactic evolution?}
\author{Francesca \surname{Matteucci}}
\runningauthor{Matteucci}
\runningtitle{Galactic Evolution} 
\institute{Department of Astronomy, University of Trieste\\
Via G.B. Tiepolo 11, 34131 Trieste, Italy}

\begin{abstract}
We are briefly introducing the most important ingredients to study 
galactic evolution. In particular the roles of star formation, 
nucleosynthesis and gas flows. Then we are discussing
the two different approaches to galactic evolution:
the
stellar population approach (chemical evolution models) and 
the hierarchical 
clustering 
scenario 
for galaxy formation. It is shown that there are still some controversial 
points in the two approaches, as evident in the brief summary of the 
discussion. 

\end{abstract}

\end{opening}

\section{Introduction}
Many parameters influence galactic evolution, in particular, we remind here
some of the most important ones:
\begin{itemize}
\item Initial conditions: closed or open model, initial chemical 
composition and gas temperature.

\item Gravity (baryonic and dark matter).

\item  Star formation history (star formation rate plus initial 
mass function).

\item  Stellar nucleosynthesis (roles of SNe).

\item  ISM processes (heating, cooling, mixing, SN feedback).

\item  Gas dynamics (enrichment of the IGM, ICM).
\end{itemize}

We intend to discuss briefly some of these items in connection with 
the constraints 
that they can impose on galaxy formation models. 
In particular, we would like to compare the predictions of current chemical 
evolution models and those of models based on the
hierarchical clustering (HC) scenario
with some relevant observational constraints relative to stellar 
populations
in ellipticals and spirals.

\section{Stellar nucleosynthesis}
Stellar nucleosynthesis is a fundamental parameter in driving galactic 
chemical
evolution by means of the stellar yields. 
Particularly important is to study the role of stars with different 
lifetimes in enriching the interstellar medium (ISM). In particular, the 
different roles played by 
SNe of different type allow us, together with the star formation history
(SFH), to interpret the 
abundance patterns in galaxies. 
What we need to know is what elements are produced by stars of different 
masses and on which timescales:
\begin{itemize}
\item Low and intermediate mass stars ($0.8 \le M/M_{\odot} \le 8$) 
are responsible for the production of $^{4}He$, C, N  and heavy
s-process elements ($A>90$). The lifetimes of these stars range from a
Hubble time to $\sim 3 \cdot 10^{7}$ years.

\item Type II SNe ($M \ge 10 M_{\odot}$)
produce mainly $\alpha$-elements 
(O, Ne, Mg, Si, S, Ca), part of Fe, light s-process (A $<$ 90) and
r-process elements.
Their lifetimes are less than $\sim 3 \cdot 10^{7}$ years. 

\item Type Ia SNe, which are believed to originate from C-deflagration 
in C-O white dwarfs,  produce mainly Fe-peak elements 
($\sim 0.6-0.7 M_{\odot}$ of Fe). 
These SNe restore their products with a time delay relative to
Type II SNe which can vary from $3 \cdot 10^{7}$ to 15 Gyr. 
Type Ia SNe are probably responsible for producing Fe in the universe 
unless the IMF is strongly biased towards massive stars. 
Their lifetimes are the same as those of low and intermediate mass stars.
\end{itemize}

Many detailed 
predictions about element production in stars of all masses are now 
available (see Matteucci 2001 for a review on this topic).

\section{The star formation history}

The SFH in a galaxy 
determines the evolution of the
ISM and its chemical composition.
The abundance patterns measured in galaxies (e.g. abundance ratios versus 
absolute abundances) are strongly influenced by their SFH.
In fact,
the time-delay between Type Ia and II SNe in restoring their nucleosynthesis 
products, together 
with the SFH determine completely the [$\alpha$/Fe] versus 
[Fe/H] distributions or any other similar abundance pattern. The influence 
of the SFH occurs via the absolute abundances whereas the 
abundance ratios are 
independent of the SFH but depend upon the assumed stellar lifetimes,
initial mass 
function (IMF) and stellar nucleosynthesis. The combination
of stellar nucleosynthesis and IMF represents the yields per stellar 
generation.  
The SFH influences the abundance ratios vs. 
abundances diagrams in the following way:
galaxies with low star formation either in bursts or continuous 
(spirals and irregulars) 
will show a short plateau in [$\alpha$/Fe] ratio 
at very low metallicities, whereas
galaxies with strong and fast SF will have a longer plateau for
[$\alpha$/Fe] (bulges, ellipticals).  
This fact can be used to infer the nature of high redshift objects, 
for example just 
by comparing the predicted tracks for galaxies of different morphological 
type with
the observed abundances in Damped Lyman-$\alpha$ objects (DLAs).
In figure 1 we show the predictions of the [$\alpha$/Fe] ratio as 
a function of [Fe/H] for different SFHs. In particular, for a spheroid 
(bulge or elliptical), the solar vicinity and a Magellanic irregular galaxy.
For comparison are shown  data points for stars and clusters in the bulge,
data for the LMC and DLAs. The data seem to be in agreement with these 
predictions. Interestingly, the data points for the DLAs seem to fall 
preferentially
on the curve relative to irregular galaxies, thus suggesting a possible 
identification of DLAs with such galaxies (see also Matteucci et al. 1997;
Calura et al. 2002).
 
\begin{figure}
\tabcapfont
\centerline{%
\begin{tabular}{c@{\hspace{1pc}}c}
\includegraphics[width=3in]{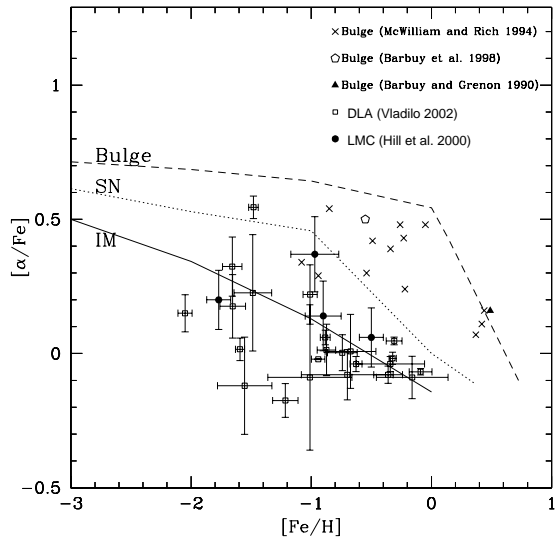} &
\end{tabular}}
\caption{The predicted [$\alpha$/Fe] ratios for three different histories of 
star formation but equal IMF and nucleosynthesis.}
\end{figure}

\section{The IMF}       
The IMF is a very crucial parameter in galactic evolution, and one should 
never try to change it unless it is absolutely necessary. 
For example, a time variable IMF could  explain the curves of Fig. 1 without 
invoking any time delay effect between SNeII and Ia in galactic enrichment.
However, this would imply a quite ``ad hoc'' variation of the IMF.
One open problem 
is to assess if the IMF is universal.
In particular,
it is not clear yet whether 
there are differences in the IMF inside galaxies and/or among galaxies. 
In our opinion there are probably IMF variations among galaxies,
whereas we cannot claim to have 
evidence for a variable (space and/or time)
IMF in the Milky Way (see Chiappini et al. 2000 for arguments 
against the variation
of the IMF).

\section{SN feedback}
How much energy is transferred from SNe and stellar winds 
into the ISM?
We would like to know the answer to this question, since it is crucial for 
understanding galactic evolution. In fact, the energy deposited by stellar 
winds and SNe into the ISM may induce gas outflows in galaxies, 
which in turn are likely to stop 
the star formation for prolonged periods.
Unfortunately, the efficiency of energy transfer is largely unknown and
always parametrized in all the  models of galactic evolution.

\section{Gas dynamics}       
Gas dynamics is crucial in determining the evolution of galaxies 
since it studies the fate of the ISM.
Gas infall and outflow are very  important parameters:
i) infall of extragalactic material is required to explain the 
G-dwarf metallicity distribution in the solar vicinity,
ii) biased infall is required to explain the formation of the Galactic 
disk and of disks in general 
(Chiappini et al. 1997; Boissier \& Prantzos 1998), iii)
galactic winds are likely to occurr in ellipticals and in 
dwarf starburst galaxies and enrich the IGM/ICM in metals
The outflows can affect the [$\alpha$/Fe] ratios inside and outside 
galaxies if the outflows are differentially enriched in heavy elements 
(Recchi et al. 2000).

\section{Monolithic vs. Hierarchical}

The modern version of the so-called monolithic model for the formation 
of elliptical galaxies
is not necessarily assuming that the galaxies form by the collapse of a 
single gas cloud but rather that ellipticals form at high redshift and
on shorter timescales 
relative to spirals, and that they are assembled out of gas and not  
of preexisting stars.
In particular, an early monolithic collapse of a gas cloud or early merging
of lumps of gas, where dissipation plays a fundamental role are equivalent
(Larson 1974;
Arimoto \&  Yoshii 1987; Matteucci \& Tornamb\`e 1987).
In the classical scenario for the formation of ellipticals,
the star formation is assumed to stop after a galactic wind develops and
all the residual gas is lost, and the galaxies are evolving passively since then.\par
The timescales required to develop a wind from  constraints on the 
properties of the stellar 
populations in local ellipticals are   
less than or equal to 1 Gyr (Pipino et al. 2002). 
Further star formation is not excluded after the wind, 
but it should not be relevant to the evolution of the stellar populations.
The winds are triggered by the energy injected 
primarily by supernova explosion and secondarily by stellar winds.
Clearly, the prescriptions for the SN feedback are a crucial ingredients in these models and are fixed by reproducing the majority of the properties of the stellar populations in ellipticals as well as the chemical enrichment in the ICM.
The appealing aspect of such models for the formation of ellipticals 
is that
they can explain the majority of the observational constraints relative 
to stellar populations: color-magnitude relations, Fundamental Plane and the
increase of the 
[$\alpha$/Fe] ratio with galactic mass (see also the detailed discussion by Peebles 2002 and references therein). \par
On the other hand,
hierarchical models of galaxy formation have the advantage of operating 
in a cosmological context. Simulations of hierarchical galaxy formation in a cold dark matter (CDM) universe are based on semi-analytical
models operating in the framework of the Press-Schechter theory.
These models predict that galaxies 
form on a wide redshift range and that massive ones form preferentially 
at late times (e.g. Kauffmann et al. 1993). In particular, 
elliptical galaxies 
should form by merger of early formed stellar systems such as spiral galaxies,
although
a burst of star formation can occur during the major merging, 
where $\sim 30\%$ of the stars can be formed (Kauffmann 1996).\par
It is clear that at least qualitatively the two approaches seem to 
predict the 
opposite trends: the hierarchical scheme predicts that spirals form before 
ellipticals which continue to assemble until recent times, as opposed to 
the monolithic model approach which predicts that ellipticals 
form on shorter 
timescales than spirals.
\par
In figure 2 we show the predictions by a model of chemical evolution 
for ellipticals
(Pipino \& Matteucci this conference) compared with the prediction by a 
model assuming the SFH derived from semi-analytical models of galaxy 
formation (Thomas 1999; Thomas et al. 2002) and with the observational data. 
As it is clearly 
shown in the figure, hierarchical models predict the opposite of what is 
observed, whereas the chemical evolution models
well reproduce the [$\alpha$/Fe] versus central velocity 
dispersion ($\sigma$) relation.
In particular, the [$\alpha$/Fe] 
trend can be obtained under the assumption that 
more massive ellipticals are older than small ellipticals, in other words 
that massive ellipticals form on shorter timescales (Matteucci 1994). 
In this case, in fact, Type Ia SNe in massive ellipticals
do not have time to substantially pollute 
the ISM before the occurrence of the galactic wind.
The high observed value of the [$\alpha$/Fe] ratio in massive ellipticals
requires that these objects formed on timescales of only few $10^{8}$ years.

\begin{figure}
\tabcapfont
\centerline{%
\begin{tabular}{c@{\hspace{1pc}}c}
\includegraphics[width=3in]{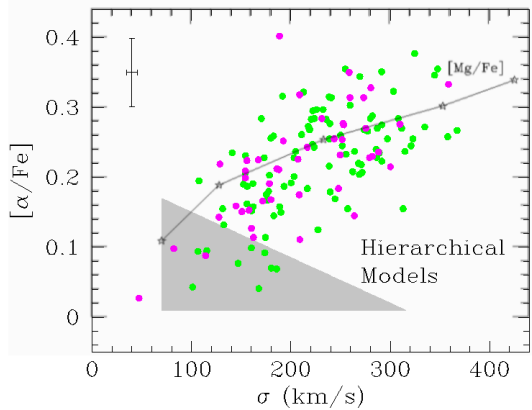} &
\end{tabular}}
\caption{Predicted and observed [$\alpha$/Fe] vs. $\sigma$ relations.
The line represents the prediction of the so-called monolithic model 
whereas the shaded area represents the predictions of a model assuming that 
ellipticals formed by merging. The figure is adapted from Thomas et al. (2002).}
\end{figure}

Disk galaxies, on the other hand, should form on much longer timescales as 
indicated by successfull models of chemical evolution of the Galaxy, which 
suggest timescales as long as 7-8 Gyr for the local disk 
(Chiappini et al. 2001).
Recent work by Boissier et al. (2001) has indicated that 
spiral galaxies in general seem to show the same tendency of ellipticals,
in the sense that massive spirals  
are older than smaller ones.
This conclusion is forced by several observational facts: in particular, 
larger
disks are redder than smaller ones, have less gas and were more active 
in the past in producing stars than lower mass disks, and the efficiency of 
star formation seems to be constant among disk galaxies.
\par
The main difference between hierarchical and ``monolithic'' approach seems 
therefore to reside in the timescales for galaxy formation and the redshift 
of formation of different objects.
>From the observational point of view there are several facts which indicate
that ellipticals formed at early times and on a short timescale, some of
these are (for a more extensive discussion see Peebles, 2002):
i) the tightness of the color-central velocity dispersion 
relation found for Virgo and Coma galaxies
(Bower et al. 1992) indicate that ellipticals in clusters have formed on 
timescales not longer than 2 Gyr. The same seems to hold for field
ellipticals, altough the timescales for these latter should be 1 Gyr older
(Bernardi et al. 1998).
ii) The thinness of the fundamental plane for ellipticals in the
same two clusters (Renzini  and Ciotti 1993) also argues in favor of a 
short period of formation for ellipticals. However, it has been suggested that
the tight relations in the fundamental plane could be due to a conspiracy 
of age and metallicity, in the sense that there is an
age-metallicity anticorrelation 
implying that
more metal rich galaxies are younger (Trager et al. 1998; 
Ferreras et al. 1999;
Trager et al. 2000).
iii) The thightness of the color-magnitude relation for 
ellipticals in clusters up to $z \sim 1$ (Kodama et al. 1998; Stanford 
et al. 1998), iv) as well as 
the modest passive evolution measured for cluster 
ellipticals at intermediate redshift (van Dokkum \& Franx 1996; 
Bender et al. 1996), argue in favor of an early formation of ellipticals.
v) The Lyman-break and SCUBA galaxies found at high redshift,
where the SFR$\sim 40- 1000 M_{\odot}yr^{-1}$ (Steidel et al. 1999;
Pettini et al. 2002: Blain et al. 1999) can be the young ellipticals
(Matteucci and Pipino, 2002).

However, there are also arguments in favor of a formation of ellipticals at 
low redshift although they seem weaker than those for the formation at high 
redshift. They are:
a) the relative large values of the $H_{\beta}$ index measured in a sample of 
nearby ellipticals which could  indicate prolonged star formation activity up 
to 2 Gyr ago (Gonzalez 1993; Faber et al. 1995; Trager et al. 1998) but see
Maraston and Thomas (2000) for an alternative explanation.
\par
b) The apparent paucity of high luminosity ellipticals 
at $z \sim 1$ compared to now (Kauffmann et al. 1996; 
Zepf, 1997;
Menanteau et al. 1998; Franceschini et al. 1998). However, 
Daddi et al. (2000) showed that the observed surface density of
extremely red objects (EROs) indicates that the bulk  of these objects 
are passively evolving ellipticals, implying that that most field 
ellipticals were fully assembled already by $z \sim 2.5$.

\section{Discussion}
In conclusion, we list some of the most 
relevant still open questions in galaxy 
evolution with the intent of stimulating a discussion.
The questions are:
Is the SFR depending on the galactic mass as suggested by observational 
evidences? 
Is the IMF universal?
Are Population III stars necessary to explain the abundance patterns 
in galaxies?
Are galactic winds from dwarf galaxies really important
for the IGM/ICM enrichment?
What are the DLAs, the Lyman-break and SCUBA galaxies?
Is inhomogeneous mixing important in the early galactic evolution?
Can the hierarchical scheme for galaxy formation reproduce 
real galaxies?
\par
At this point the discussion started and was concentrated on the last 
of the open questions listed 
above, in particular on the comparison between hierarchical and monolithic
models for galaxy formation.
\par
Daniel Thomas started the discussion by showing 
a summary of observations which 
indicate some problems connected with hierarchical clustering models for 
the formation of galaxies. Some of them were 
addressed as ``classical arguments'', whereas others
were addressed as ``new arguments''.
The classical arguments are:
missing satellites, angular momentum, cuspy cores.
The new arguments are:
the color-magnitude relation, the Fundamental Plane,
[$\alpha$/Fe]-sigma relation,
anti-hierarchical ages of ellipticals, ages of bulges,
chemical homogeneity of thick disk,
ULIRGs (ultra luminous infrared galaxies) 
do not evolve onto the Fundamental Plane,
Scuba sources look like ellipticals at high redshift with very intense SFR,
boxiness requires E--E mergers.
\par
Rosa Dominguez argued that in her simulations of galaxy formation 
in the $\Lambda$CDM scenario
ellipticals can form at early times and the massive ellipticals 
on shorter timescales than the smaller ellipticals.
Silk argued that it is not correct to oppose the two scenarios and that the 
hierachical clustering scenario is essentially a model for the dark matter.
Matthias Steinmetz said that quantitative calculations to fit the 
[$\alpha$/Fe] ratios with hierarchical models have never been performed
because of the difficulty of obtaining a good numerical resolution, but that 
certainly galaxies did not form out of a collapsing sphere!
(we all agreed on that!).  However, he said that not necessarily HC models would predict the opposite of what is observed. In fact, he pointed out that although massive ellipticals would form later, this can be different in cluster environments where
massive galaxies would form faster. As most of ellipticals are in clusters this could give rise to timescales in agreement with the observed [$\alpha$/Fe] versus
velocity dispersion relation.Then he said 
that the spirals which merge to form the ellipticals are different from the
present day spirals and that not all ellipticals form from spiral mergers.
Cristina Chiappini then replied that if there is a difference in the 
formation timescales for ellipticals in clusters and in the field, then 
we should see different chemical properties and this is not observed as 
shown by Bernardi et al. (1998).
Kobayashi said that in the $\Lambda$CDM scenario is still possible to have 
monolithic ellipticals. She also 
suggested that the presence of abundance gradients in ellipticals argue 
against the formation of ellipticals by mergers, whereas,  on the contrary, 
ellipticals with no gradients can derive from mergers.
Steven Shore 
recommended to include mergers in chemical evolution models before claiming 
that the real scenario cannot be hierarchical. He suggested in particular 
that chemical evolution models could test the infall timescales coming 
from simulations of galaxy formation.
Cristina Chiappini replied that in the specific case of the Milky Way, 
the infall law suggested by Sommer-Larsen et al. (2002) in his HC model
seems to be in good agreement with the infall law used in chemical 
evolution models which are able to reproduce the majority of the 
observational constraints.
Hensler wondered whether dissipational mergers of spirals may create 
ellipticals in a way similar to monolithic collapse.
However, Burkert said that his models for elliptical formation do need 
major mergers.
\par
What emerged from this discussion was that in the future 
we need to combine detailed chemical evolution with semi-analytical models
of galaxy formation in order to try to understand the many still 
open questions concerning 
galaxy formation and evolution.
\par
In concluding, I would like to thank all of those who partecipated to the 
discussion and apologize if I missed other contributions to the discussion,
since everybody has contributed to a very lively discussion.

\end{article}
\end{document}